\documentclass[12pt]{iopart}

\usepackage{graphicx,times,iopams}
\usepackage{amssymb,mathrsfs,hyperref}

\hypersetup{
colorlinks=true,
linkcolor=black,          
citecolor=blue,           
filecolor=magenta,        
urlcolor=cyan
}

\newcommand{\ket}[1]{\left| #1 \right\rangle}
\newcommand{\bra}[1]{\left\langle #1 \right|}

\newcommand{\elph}{{\rm el-ph}}
\newcommand{\ph}{{\rm ph}}

\newcommand{\hc}{{\rm h.c.}}
\newcommand{\arccosh}{{\rm arccosh}}
\renewcommand{\Re}{{\rm Re\ }}
\renewcommand{\Im}{{\rm Im\ }}
\newcommand{\Hh}{\mathcal{H}}

\newcommand{\tc}{\tilde{c}}

\renewcommand{\L}{\mathcal{L}}

\begin{document}

\title{Microwave-controlled coupling of Majorana bound states}
\author{Thomas L. Schmidt, Andreas Nunnenkamp and Christoph Bruder}
\address{Department of Physics, University of Basel, Klingelbergstrasse 82, CH-4056 Basel, Switzerland}
\ead{thomas@thoschmidt.de}

\begin{abstract}
We propose microwave-controlled rotations for qubits realized as
Majorana bound states. To this end we study an inhomogeneous Kitaev
chain in a microwave cavity. The chain consists of two topologically
nontrivial regions separated by a topologically trivial, gapped
region.  The Majorana bound states at the interfaces between the
left (right) regions and the central region are coupled, and their
energies are split by virtual cotunneling processes. The amplitude
for these cotunneling processes decreases exponentially in the
number of sites of the gapped region, and the decay length diverges
as the gap of the topologically trivial region closes. We
demonstrate that microwave radiation can exponentially enhance the
coupling between the Majorana bound states, both for classical and quantized electric fields. By solving the appropriate
Liouville equation numerically we show that microwaves can drive
Rabi oscillations in the Majorana sector. Our model emerges as an
effective description for a topological semiconductor nanowire in a
microwave cavity. Thus, our proposal provides an
experimentally feasible way to obtain full single-qubit control
necessary for universal quantum computation with Majorana qubits.
\end{abstract}

\maketitle

\section{Introduction}

Majorana bound states (MBS) \cite{kitaev01} are currently a strong
focus of research in the condensed-matter community
\cite{beenakker11,leijnse12,alicea12}. Semiconductor nanowires with
strong Rashba spin-orbit coupling have emerged as a promising platform
to host MBS. Following theoretical proposals \cite{sato09, lutchyn10, oreg10}
first experimental signatures of MBS have recently been reported
\cite{mourik12, rokhinson12, deng12, das12}.

On the one hand, MBS are fascinating quantum systems in their own
right. Despite the fact that Majorana fermions were predicted theoretically 70 years ago \cite{majorana37}, it remains unclear to this date whether they exist as elementary particles. This means MBS may be their closest living relative. On the other hand, MBS could become useful for quantum computing
because braiding MBS makes it possible to implement topologically-protected qubit operations which are immune against certain types of noise \cite{nayak08}.
However, these braiding operations do not form a universal set of gates
needed for quantum computation, so they have to be supplemented by
other gates which are not topologically protected.
The circuit quantum-electrodynamics (circuit-QED) architecture \cite{Schoelkopf2008}
offers a controlled and well-developed toolbox in the microwave domain and is thus an ideal candidate to complement the topologically-protected braiding
operations. Moreover, hybrid structures involving semiconductor nanostructures and microwave cavities have already been realized experimentally \cite{frey12}. Coupling MBS of a Kitaev chain to a microwave strip-line resonator has recently been studied theoretically by Trif and Tserkovnyak \cite{trif12}.

We propose to use microwaves to control the coupling between two
MBS, which is potentially relevant in the context of quantum
computation. We will show that photon-assisted tunneling has a strong impact on the coupling between adjacent MBS which are separated by a short topologically trivial region. In Ref.~\cite{schmidt13a}, we presented the main idea using a minimal model containing only two MBS coupled via a gapped region with quadratic spectrum. We used perturbation theory for weak electron-photon coupling to derive an analytic result for the system dynamics, and found that a microwave field can induce Rabi oscillation between adjacent MBS. In this paper, we extend that analysis by considering a more realistic model and treating the coupling exactly. We explain how coupling between MBS and a microwave cavity field emerges in a semiconductor nanowire hosting MBS, and map the corresponding Hamiltonian onto a Kitaev chain coupled to a cavity. We solve this model numerically and obtain a solution valid at arbitrary electron-photon coupling strength.

The remainder of this paper is organized as
follows. In Section \ref{sec:kitaev} we introduce the Hamiltonian for
an inhomogeneous Kitaev chain consisting of two topologically
nontrivial regions separated by a topologically trivial, gapped
region. In this situation it is well-known that MBS form at the
interfaces between the topologically different phases \cite{kitaev01}.
We demonstrate that virtual cotunneling processes mediated by the
gapped region couple the MBS. The amplitude
for these processes decreases exponentially in the length of the
gapped region, and the decay length diverges as the gap in the
topologically trivial region closes.

The Kitaev model is known to be an effective description for a topological nanowire
\cite{alicea11}. In Section \ref{sec:elph} we study the coupling of such a nanowire to the microwave field inside a cavity. We find that the cavity field
gives rise to a modulation of the hopping matrix element between
neighbouring lattice sites. This enables us to control the properties
of the chain with the microwave field.

In Section \ref{sec:kitaev_micro} we discuss a Kitaev model coupled to a
microwave field. In the first part of that section, we assume that the cavity field can be described classically and investigate the dynamics of the driven system. In the second part, we solve the full quantum master equation for a
damped, quantized photon field. In both cases we
find that microwaves can exponentially enhance Rabi oscillations in the Majorana
sector. They can thus supplement topologically-protected braiding
operations to gain full single-qubit control which is a necessary step
toward a universal set of quantum gates for Majorana qubits.

\section{Inhomogeneous Kitaev chain}
\label{sec:kitaev}
To describe a one-dimensional wire which can be brought into either a
topologically trivial or topologically nontrivial phase by tuning the
system parameters, we shall use the Hamiltonian of the Kitaev chain
\cite{kitaev01}. This model captures qualitatively many features of
more realistic 1D models \cite{lutchyn10,oreg10}. In fact, it has
been shown that there exists an approximate mapping between these more realistic 1D
models and the Kitaev chain Hamiltonian, so their low-energy degrees
of freedom are identical \cite{alicea11}. We shall extend this mapping to incorporate
electron-photon coupling in Sec.~\ref{sec:elph}.

We start with a brief review of some essential properties of an
\emph{inhomogeneous} Kitaev chain, containing two topologically
nontrivial regions, separated by a short topologically trivial, gapped
region. We shall derive the effective low-energy Hamiltonian describing the overlap between the two MBS adjacent to the central region, and show that such a Hamiltonian can in principle be used to implement single-qubit rotations.
On a very general level, this system can be modeled as a
Kitaev chain consisting of $N$ sites with position-dependent
parameters,
\begin{equation}\label{eq:kitaev}
    H_K = - \sum_{n = 1}^N \mu_n c^\dag_n c_n
    - \frac{1}{2} \sum_{n=1}^{N-1} \left( t_n c^\dag_n c_{n+1} + \Delta_n c_n c_{n+1} + \hc \right).
\end{equation}
Here, $\mu_n$ denote the onsite energies, $t_n$ are the hopping
amplitudes, and $\Delta_n$ are the $p$-wave pairing strengths. The
operator $c_n$ ($c_n^\dag$) annihilates (creates) a spinless fermion
at lattice site $n$. This bilinear Hamiltonian can be diagonalized
exactly for arbitrary parameters and will be used below for numerical
results. To obtain simple analytical expressions, it is
convenient to select the simplest set of parameters which
generates the desired topological phases. We therefore assume
Eq.~(\ref{eq:kitaev}) to be of the form
\begin{equation}\label{eq:kitaev_simp}
    H_K = H_L + H_C + H_R + H_{LC} + H_{RC},
\end{equation}
where the $N$-site chain is split into three parts: $H_L$ and $H_R$
describe the left ($n \leq m_1$) and right ($n \geq m_2$) parts of the
chain, and $H_C$ the $N_C = m_2 - m_1 - 1$ sites in the central part
($m_1+1 \leq n \leq m_2-1$). The sections are coupled by $H_{LC}$, which
connects sites $m_1$ and $m_1+1$, and $H_{RC}$, which connects sites
$m_2-1$ and $m_2$.

The left and right regions are supposed to be in the topologically
nontrivial phase, i.e., we assume $\mu_n = 0$ and constant $\Delta_n =
t_n > 0$ in these regions. Then, the Hamiltonians $H_{L,R}$ become diagonal in a
basis of nonlocal Dirac fermions, $d_n = \Im c_{n+1} + i \Re c_n$,
\begin{equation}
    H_L = t_L \sum_{n = 1}^{m_1 - 1} d^\dag_n d_n,
\qquad  H_R = t_R \sum_{n = m_2}^{N - 1} d^\dag_n d_n.
\end{equation}
In the central part of the chain, we choose the parameters as $\mu_n =
\mu_C$, $t_n = t_C > 0$, and $\Delta_n = 0$. Without loss of generality, we
assume $\mu_C > 0$. Since this parameter choice makes the central
region topologically trivial, we retain the basis of local Dirac
fermions $c_n$,
\begin{equation}
    H_C = - \mu_C \sum_{n=m_1+1}^{m_2-1} c^\dag_n c_n -
\frac{t_C}{2} \sum_{n=m_1+1}^{m_2-2} \left(  c^\dag_n c_{n+1} + \hc \right).
\end{equation}
To diagonalize $H_C$, it is convenient to extend the central region to
a large number of sites $N_\infty \gg N_C$ and then impose periodic
boundary conditions. We shall discuss the quality of this
approximation by comparing it to numerical results below. Hence, we
introduce the operators $\tc_n$, where $\tc_n = \tc_{n +
  N_\infty}$. They are defined to coincide with the original operators
$c_n$ in the central region: $\tc_n = c_n$ for $m_1 + 1 \leq n \leq
m_2 -1$.  Then, we can diagonalize $H_C$ in momentum space,
\begin{equation}
    H_C = \sum_k \epsilon(k) \tc_k^\dag \tc_k,
\end{equation}
where $\epsilon(k) = - \mu_C - t_C \cos(a_0 k)$. We introduced the
lattice spacing $a_0$, and the momentum $k$ is in the first Brillouin
zone, $k \in [-\pi/a_0, \pi/a_0]$, and is quantized in units of
$2\pi/(a_0 N_\infty)$. The operators in momentum space are defined by
$\tc_k = N_{\infty}^{-1/2} \sum_{n=1}^{N_\infty} e^{-i k a_0 n}
\tc_n$.

\begin{figure}[t]
\includegraphics[width=10cm]{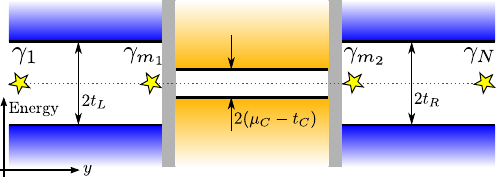}
\centering
\caption{Band structure of the inhomogeneous Kitaev chain along the $y$ axis: the left and right chain
  segments are in the topologically nontrivial phase and host
  Majorana bound states (MBS) at their edges. The central region is gapped and
  topologically trivial. Stars denote the positions of
  zero-energy MBS.}
    \label{fig:Schematic}
\end{figure}

If the three segments are not coupled ($H_{LC} = H_{RC} = 0$), our
choice of parameters entails that the left (right) segment is
gapped due to the superconducting pairing and the width of the gap is
$2 t_L$ ($2 t_R$). The uncoupled Hamiltonian has a four-fold degenerate ground state as there are two MBS, $\gamma_1$ and $\gamma_{m_1}$, at the ends of the left chain and two MBS, $\gamma_{m_2}$ and $\gamma_N$, at the ends of the right chain.
In terms of the electron operators, these are given by,
\begin{eqnarray}\label{eq:gamma}
    \gamma_1 &= 2 \Im c_1 = -i (c_1 - c_1^\dag), \nonumber \\
    \gamma_{m_1} &= 2 \Re c_{m_1} = c_{m_1} + c^\dag_{m_1}, \nonumber \\
    \gamma_{m_2} &= 2\Im c_{m_2} = -i (c_{m_2} - c_{m_2}^\dag), \nonumber \\
    \gamma_N &= 2 \Re c_N = c_{N} + c^\dag_N.
\end{eqnarray}
Now, we introduce tunneling between the side chains and the central
chain. Because there are no terms beyond nearest-neighbor terms in
Eq.~(\ref{eq:kitaev}), only the Majorana states $\gamma_{m_1}$ and
$\gamma_{m_2}$ are coupled to the central chain \cite{bolech07},
\begin{eqnarray}\label{eq:H_LRC}
    H_{LC} &= - \frac{t_{LC}}{2} \gamma_{m_1} (\tc_{m_1 + 1} - \tc^\dag_{m_1 +1}), \nonumber \\
    H_{RC} &= - \frac{i t_{RC}}{2} (\tc_{m_2 - 1} + \tc^\dag_{m_2 - 1}) \gamma_{m_2}.
\end{eqnarray}
Note that for our choice of parameters the Dirac fermions $d_n$ of the
left and right chains are not coupled to the remaining
system. Therefore, we can discard the operators $H_{L,R}$ and use
\begin{equation}\label{eq:Hsimple}
    H_0 = H_C + H_{LC} + H_{RC}.
\end{equation}
The Hamiltonian (\ref{eq:Hsimple}) can easily be solved analytically. We
are particularly interested in the coupling between $\gamma_{m_1}$ and
$\gamma_{m_2}$, mediated by the central chain. For $\mu_C < t_C$, the
central chain is gapless and real electrons and holes can tunnel
to and from the MBS. In this regime, the overlap of the MBS turns out
to be very sensitive to system parameters, and the boundary conditions
of the central region become important. The MBS self-energy acquires
an imaginary part, indicating a level broadening and thus a finite
lifetime of the MBS. In the continuum limit, the effect of the
metallic central region thus resembles that of a fermionic bath which
has been investigated in detail in Ref.~\cite{budich12a}.

\begin{figure}[t]
    \includegraphics[width=10cm]{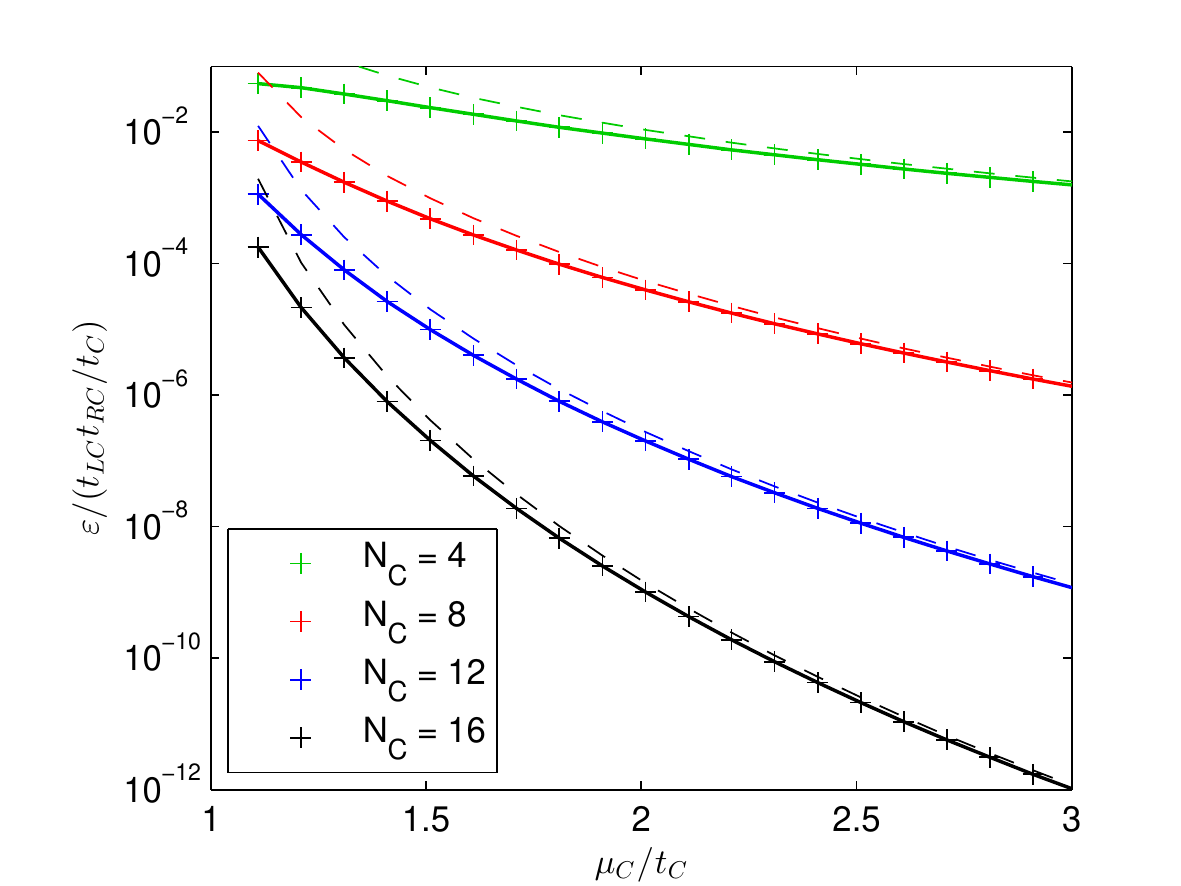}
    \centering
    \caption{Splitting of the MBS as a function of chemical potential $\mu_C$ for different numbers of lattice sites $N_C$ in the central region. The crosses indicate the numerically exact solution of the inhomogeneous Kitaev chain, see Eq.~(\ref{eq:kitaev}). The solid line represents the splitting calculated numerically using periodic boundary conditions. The dashed line shows the approximate analytic expression (\ref{eq:shift_dyson}) which is valid in the limit $\mu_C \gg t_C$.}
    \label{fig:Splitting}
\end{figure}

On the other hand, for $\mu_C > t_C$, the central chain has a gap of
width $2 ( \mu_C - t_C)$. The corresponding band structure (for uncoupled chains) is shown in Fig.~\ref{fig:Schematic}. In this regime,
electron cotunneling is the only process leading to a nonzero overlap
of the Majorana levels $\gamma_{m_1}$ and $\gamma_{m_2}$. This overlap
causes a finite level splitting of the MBS, which for $\mu_C
\gg t_C$ becomes
\begin{equation}\label{eq:shift_dyson}
    \varepsilon =  t_{LC} t_{RC} \frac{e^{-(N_C-1) \arccosh(\mu_C/t_C)}}{t_C\sqrt{(\mu_C/t_C)^2 - 1}}.
\end{equation}
Importantly, this splitting decays exponentially with increasing
length $L_C = a_0N_C$ of the central region, which is the reason for the topological protection of the MBS. The corresponding decay length
$\xi = a_0/\arccosh(\mu_C/t_C)$ diverges as $(\mu_C/t_C - 1)^{-1/2}$
for $\mu_C \to t_C$. The level splitting (\ref{eq:shift_dyson}) can also
be found from the exact numerical solution of the full Hamiltonian
(\ref{eq:kitaev}). A comparison between the exact numerical result and
the approximation (\ref{eq:shift_dyson}) is shown in
Fig.~\ref{fig:Splitting}.

In Ref.~\cite{schmidt13a}, we calculated the level splitting for a
minimal model containing two MBS coupled via a gapped region with
quadratic spectrum $\epsilon(k) = k^2/(2M) - \mu_0$. To
compare the energy splitting (\ref{eq:shift_dyson}) with that model,
we need to take the continuum limit of the Kitaev chain by sending the
lattice constant $a_0 \to 0$, while keeping the total length $L_C$ constant. Let us assume that $\mu_C \gtrsim t_C$. Near the
bottom of the band the spectrum of the uncoupled central Kitaev chain
is quadratic with effective mass $M = 1/(a_0^2 t_C)$ and chemical
potential $\mu_0 = t_C-\mu_C$. Then, one finds for small $|\mu_0| \ll
t_C$,
\begin{equation}
    \varepsilon \approx a_0 t_{LC} t_{RC} \frac{e^{-L_C/\xi}}{2 |\mu_0| \xi}
\end{equation}
where $\xi = (2 M |\mu_0|)^{-1/2}$ is the decay length of the MBS into
the gapped system.

If we focus on the gapped regime ($\mu_C > t_C$), we can derive a simple
effective low-energy theory of the coupled Majorana modes
$\gamma_{m_1}$ and $\gamma_{m_2}$ by integrating out the central
region. This leads to an effective retarded interaction between
$\gamma_{m_1}$ and $\gamma_{m_2}$. In the low-energy (long-time) limit
at energy scales small compared to $(\mu_C^2 - t_C^2)/\mu_C$, we can
neglect the retardation and obtain the effective low-energy
Hamiltonian
\begin{equation}\label{eq:H0_low_energy}
    H_{\mathrm{eff}} = \frac{i \varepsilon}{2} \gamma_{m_1} \gamma_{m_2}.
\end{equation}
For the purposes of topologically protected quantum computing, a logical qubit should be encoded into four MBS \cite{alicea12}, which can in turn be combined into two Dirac fermions, e.g., $\psi_L = (\gamma_1 + i\gamma_{m_1})/2$ and $\psi_R = (\gamma_{m_2} + i \gamma_N)/2$. Since the topological protection relies on a conserved fermion parity, the computational basis should contain states with equal parity, e.g., the odd-parity states
\begin{eqnarray}\label{eq:eom_qubit}
    \ket{\downarrow} &=& \psi_L^\dag \ket{0}, \nonumber \\
    \ket{\uparrow} &=&  \psi_R^\dag \ket{0}
\end{eqnarray}
where $\ket{0}$ denotes the ground state of the system which is annihilated by $\psi_R$ and $\psi_L$. Braiding the MBS $\gamma_1$ and $\gamma_{m_1}$ ($\gamma_{m_2}$ and $\gamma_N$) corresponds to the unitary transformations $U_{1,2} = \exp\left( \pm \frac{i \pi}{4} \sigma_z \right)$, respectively, where $\sigma_z$ is a Pauli matrix in the basis $\{ \ket{\uparrow}, \ket{\downarrow} \}$. Similarly, braiding $\gamma_{m_1}$ and $\gamma_{m_2}$ corresponds to $U_3 = \exp\left(i \pi \sigma_x/4 \right)$. Clearly, these operations are insufficient to reach arbitrary points on the Bloch sphere \cite{nayak08}. In order to perform arbitrary single-qubit rotations, one needs to supplement them by topologically unprotected single-qubit gates. For instance, if the MBS $\gamma_{m_1}$ and $\gamma_{m_2}$ are subject to a coupling Hamiltonian (\ref{eq:H0_low_energy}) for a certain time $t$, the qubit state will be rotated by the unitary transformation
\begin{equation}\label{eq:U}
    U_\varepsilon(t) = \exp\left( - \frac{i \varepsilon t}{2} \sigma_x \right).
\end{equation}
It is easy to show that arbitrary points on the Bloch sphere can now be reached by combining the operations $U_\varepsilon(t)$ with a single braiding operation, e.g., $U_1$ \cite{schmidt13a}. Let us stress again that the operation $U_\varepsilon(t)$ is not topologically protected. In the following, we shall demonstrate that the coupling Hamiltonian (\ref{eq:H0_low_energy}) can be engineered by using the interaction of MBS with a microwave cavity field. A protocol to braid MBS using $T$-junctions of semiconductor nanowires was presented in Ref.~\cite{alicea11}. By combining this proposal with a microwave cavity, as depicted in Fig.~\ref{fig:SetupCavity}, it is thus possible to realize arbitrary single-qubit operations.

\section{Coupling of Majorana bound states to photons}\label{sec:elph}

In the proposed solid-state devices \cite{sato09,lutchyn10,oreg10},
MBS exist as quasiparticles consisting of an equal-weight
superposition of a particle and a hole. Therefore, they may interact
with photons despite the fact that they are on average chargeless. In
order to gain further insight into this coupling, we shall use the
Hamiltonian of Ref.~\cite{oreg10}, and couple it to an electric field
using the minimal-coupling substitution $\vec{p} \to \vec{p} -
e\vec{A}$, where $\vec{p}$ is the momentum operator, $e$ is the elementary charge, and $\vec{A}$ is the
vector potential.

The model Hamiltonian for a semiconductor nanowire along the $y$-axis reads
\begin{eqnarray}\label{eq:H_Oreg}
    H_{\mathrm{nw}} &=& \int dy \Bigg\{ \sum_{\sigma = \uparrow,\downarrow} \psi^\dag_\sigma(y) \left[ - \frac{1}{2M} \frac{\partial^2}{\partial y^2} - \mu_0 - i u \sigma \frac{\partial}{\partial y} \right] \psi_\sigma(y) \nonumber \\
    &+& \sum_{\sigma = \uparrow,\downarrow} B \psi^\dag_{\sigma}(y) \psi_{-\sigma}(y) + \left[ \Delta_0 \psi^\dag_\uparrow(y) \psi^\dag_\downarrow(y) + \hc \right] \Bigg\}.
\end{eqnarray}
Here, $\mu_0$ denotes the chemical potential, and $u$ is the strength
of the Rashba spin-orbit coupling pointing in the $z$ direction. $B$
is a perpendicular magnetic field in the $x$ direction. Finally, the
induced $s$-wave pairing strength is denoted by $\Delta_0$. It has
been shown that, assuming $B > \Delta_0$, this model contains MBS at the
edges for $|\mu_0| < \sqrt{B^2 - \Delta_0^2}$, whereas it is in the
topologically trivial phase for $|\mu_0| > \sqrt{B^2 - \Delta_0^2}$.

\begin{figure}[t]
\includegraphics[width=10cm]{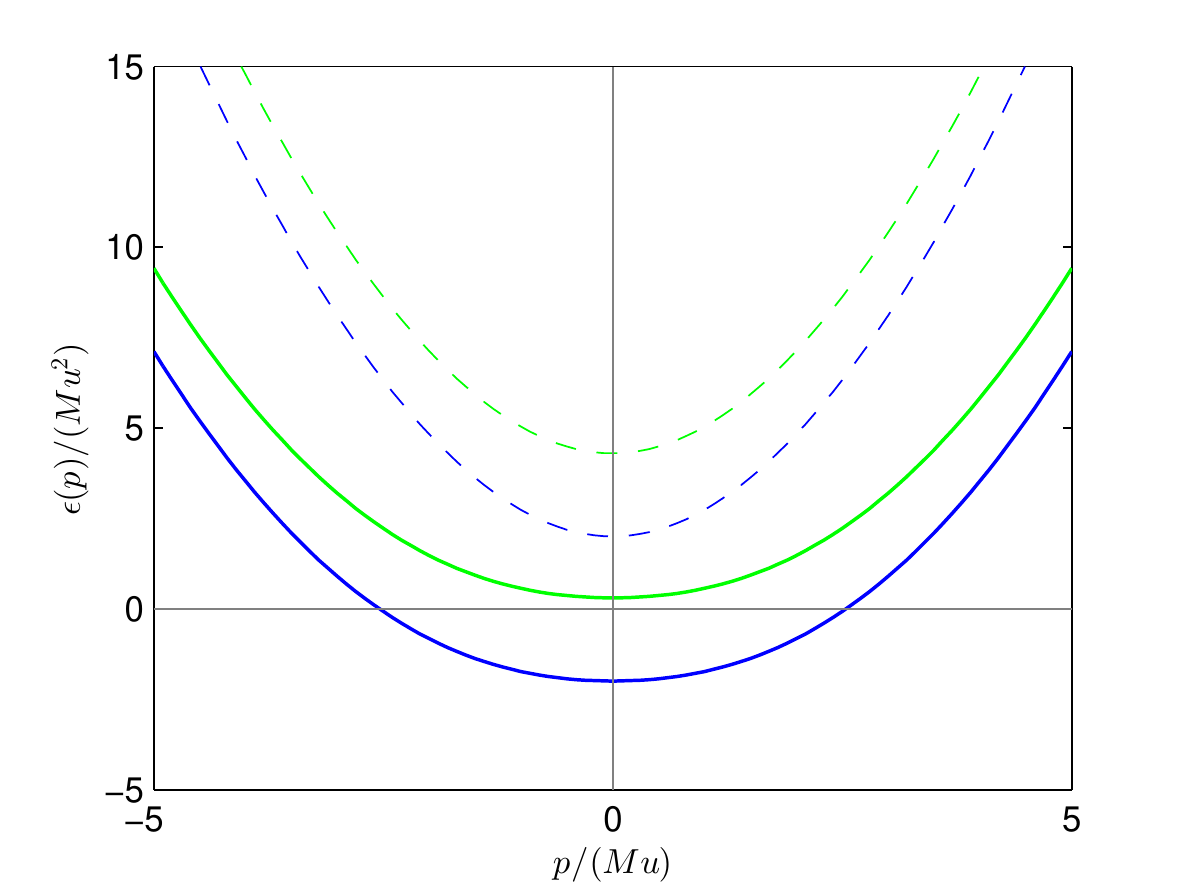}
\centering
\caption{Single-particle spectrum corresponding to
  Eq.~(\ref{eq:H_Oreg}) for $\Delta_0 = 0$. Blue lines are for the
  topologically nontrivial phase ($\mu_0 = 0$, $B = 2 M u^2$), green
  lines are for the trivial phase ($\mu_0 = -2.3 M u^2$, $B = 2 M
  u^2$). For the mapping onto a spinless Kitaev chain, the
  higher-energy parts of the spectrum (dashed lines) are neglected.}
\label{fig:epsilonPlot}
\end{figure}

Mapping Eq.~(\ref{eq:H_Oreg}) on a Kitaev chain becomes most
transparent in the regime $B \gg \Delta_0, M u^2$. The band structures
in the topologically trivial (nontrivial) phases are depicted in
Fig.~\ref{fig:epsilonPlot}. In the nontrivial regime, we can choose
$\mu_0 = 0$ and the band gap is proportional to $\Delta_0 u/B$. In the trivial regime, on the other hand, we choose
$-\mu_0 \gtrsim B$ and the band gap is $|\mu_0| - B$. One obtains a
low-energy theory by retaining the lower branches $\psi_- \approx (\psi_\uparrow - \psi_\downarrow)/\sqrt{2}$ of both
spectra \cite{alicea11}. In the nontrivial region, the result is a spinless $p$-wave
superconductor. On the other hand, the trivial region becomes a spin-polarized
electron system. Both systems may be described by Kitaev chains in the
respective topological phases. For a fixed lattice spacing $a_0$, the
parameters of the homogeneous Kitaev chain
\begin{equation}
H_K = - \sum_{n = 1}^N \mu c^\dag_n c_n
    - \frac{1}{2} \sum_{n=1}^{N-1} ( t c^\dag_n c_{n+1} + \Delta c_n
    c_{n+1}
+ \hc )
\end{equation}
are related to those of the model
Hamiltonian $H_{\mathrm{nw}}$ as follows,
\begin{equation}
    \mu = \mu_0 + B - \frac{1}{M a_0^2},\quad
    t = \frac{1}{M a_0^2},\quad
    \Delta = \frac{\Delta_0 u}{B a_0}\:.
\end{equation}
For our proposed coupling scheme, the band gap in the nontrivial
region should ideally be larger than in the trivial regions, so we will assume
that $|\mu_0|, B \gg \Delta_0$, whereas $|\mu_0| - B \ll \Delta_0$.

\begin{figure}[t]
\includegraphics[width=10cm]{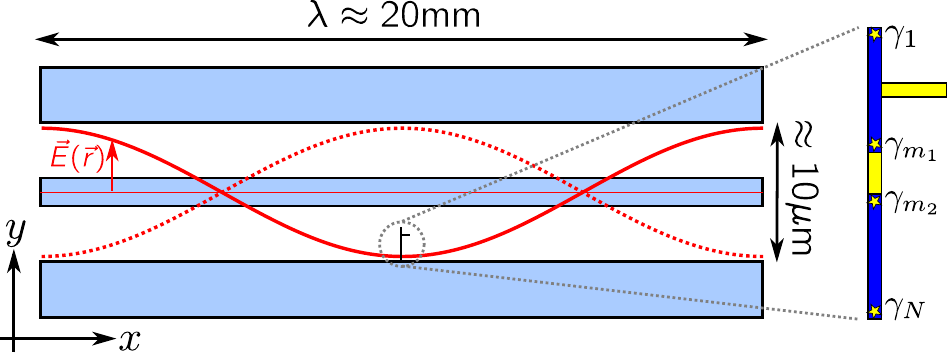}
\centering
\caption{The microwave cavity is realized as a superconducting stripline. The nanowire hosting MBS is located close to one of the maxima of the electric field $\vec{E}(\vec{r})$ and is parallel to $\vec{E}$. A $T$-junction in the nanowire allows braiding of two MBS, whereas the microwave field allows Rabi oscillations between the two other MBS.}
    \label{fig:SetupCavity}
\end{figure}

The coupling to a vector potential $\vec{A}(\vec{r})$ in a
microwave cavity can now be investigated using the minimal coupling
substitution $-i \partial/\partial \vec{r} \to -i \partial/\partial
\vec{r} - e\vec{A}(\vec{r})$ in the Hamiltonian (\ref{eq:H_Oreg}). We
assume that the cavity electric field $\vec{E} = - d\vec{A}/dt$ is
oriented along the wire ($y$) axis, i.e., the wire axis is
perpendicular to the cavity axis, see Fig.~\ref{fig:SetupCavity}. Moreover, we assume that
$\vec{E}(\vec{r})$ is spatially constant along the length of the wire,
which is a good approximation for current experimental setups
\cite{mourik12} and typical microwave wavelengths. In that case,
$A_y = (E_{\mathrm{rms}}/\Omega)  (a + a^\dag)$, where $E_{\mathrm{rms}}$ is the root mean square of the cavity electric field, $a$ ($a^\dag$) is the
annihilation (creation) operator for the cavity mode, and $\Omega$ is its frequency. Since $A_y$ is
position-independent, it commutes with the momentum operator, so the
electron-photon coupling Hamiltonian becomes
\begin{equation}\label{eq:map_Helph}
    H_{\rm el-ph}
=
    \frac{i e A_y}{M} \int dy \psi_-^\dag(y) \partial_y \psi_-(y).
\end{equation}
We can again discretize the spatial integral. For a lattice spacing $a_0$, the result reads
\begin{equation}\label{eq:map_Helph_lattice}
    H_{\rm el-ph} = \frac{e A_y}{2 M a_0} \sum_n \left( i c_n^\dag c_{n+1} + \hc \right) \propto i (a + a^\dag)  \sum_n c_n^\dag c_{n+1} + \hc
\end{equation}
In conclusion, an electric field gives rise to a change in the hopping matrix element between neighboring lattice sites. The expression (\ref{eq:map_Helph_lattice}) holds in both the topological trivial and nontrivial regimes. We expect the same type of coupling also at interfaces between different phases, and the coupling constant will then depend on the overlap between the wavefunctions in both regions and on their spin structure.

\section{Inhomogeneous Kitaev chain in a microwave cavity}
\label{sec:kitaev_micro}

Let us assume that the inhomogeneous Kitaev chain is brought into a
driven microwave cavity \cite{trif12}. We consider one cavity mode with frequency
$\Omega$ and assume that the cavity is driven with a frequency
$\Omega_L$,
\begin{equation}\label{eq:Hcav}
    H_{\rm cav}(\tau) = \Omega a^\dag a + a \varphi e^{i \Omega_L \tau} +
a^\dag \varphi^* e^{-i \Omega_L \tau} + H_{\rm cav,d},
\end{equation}
where $\tau$ is the time, $a$ is the bosonic operator of the cavity
mode, $\varphi$ represents amplitude and phase of the drive, and
$H_{\rm cav,d}$ contains damping terms, which produce a nonzero line
width $\kappa$.

We assume that the wavelength of the field $\lambda = c/\Omega$ is much longer than the Kitaev chain, so we can treat the field as constant along the Kitaev chain. The cavity field acts on the electrons forming the Kitaev chain. Thus, we use the Hamiltonian (\ref{eq:kitaev}) and supplement it by electron-photon coupling terms (\ref{eq:map_Helph_lattice}).

For the numerical solution of the system, we represent the Kitaev Hamiltonian~(\ref{eq:kitaev}) as $H_K = \frac{1}{2} A^\dag \Hh_K A$, where $A^\dag = ( c_1^\dag, \ldots, c_N^\dag, c_1, \ldots, c_N )$ and $\Hh_K$ is a complex $2N \times 2N$ matrix. In the absence of coupling between the three regions, the matrix $\Hh_K$ has a fourfold degenerate eigenvalue at zero energy reflecting the four MBS (\ref{eq:gamma}). The corresponding basis of the zero-energy eigenspace is,
\begin{eqnarray}\label{eq:psi_j}
    \ket{\psi_1} = \frac{i}{\sqrt{2}} \left( \ket{1} - \ket{1 + N} \right),\quad &
    \ket{\psi_{m_1}} = \frac{1}{\sqrt{2}} \left( \ket{m_1} + \ket{m_1 + N} \right)  \nonumber \\
    \ket{\psi_{N}} = \frac{1}{\sqrt{2}} \left( \ket{N} + \ket{2 N} \right),\quad &
    \ket{\psi_{m_2}} = \frac{i}{\sqrt{2}} \left( \ket{m_2} - \ket{m_2 + N} \right)
\end{eqnarray}
where $\ket{n}$ is the vector with components $\ket{n}_j = \delta_{nj}$ for $n \in \{1, \ldots, 2 N\}$. For the numerical simulation, we shall assume that the system is initially prepared in the state $\ket{\psi_{m_1}}$. The time evolution
then leads to Rabi oscillations between the states $\ket{\psi_{m_1}}$ and $\ket{\psi_{m_2}}$.

To apply the proposed coupling mechanism for qubit rotations, the gap in the topologically nontrivial left and right regions should exceed the gap in the topologically trivial central region, as shown in Fig.~\ref{fig:Schematic}. Moreover, the photon frequency $\Omega$ should be slightly below the gap in the central region, and the left and right regions much longer than the central region. This choice of parameters ensures that the two MBS in the left region ($\gamma_1$ and $\gamma_{m_1}$) and the right region ($\gamma_{m_2}$ and $\gamma_N$) remain unaffected by the photon field, whereas $\gamma_{m_1}$ and $\gamma_{m_2}$ will be coupled. In our numerical simulations, we choose $\mu_L = \mu_R = 0$. In this case, the MBS within the left and right regions are mutually uncoupled, so it is sufficient to consider small lengths $N_{L,R}$.

\subsection{Classical microwave field}\label{sec:kitaev_classical}

Let us now first consider the case where the microwave field can be treated classically. Moving into a rotating frame with the drive frequency $\Omega_L$, the classical approximation corresponds to replacing the quantum mechanical operators $a,a^\dag$ by $a \rightarrow \sqrt{n_\ph} e^{-i \Omega_L \tau}$. The Hamiltonian is then independent of the cavity resonance frequency $\Omega$ and reads
\begin{equation}\label{eq:Hcl}
    H = H_K - 2 \beta \sqrt{n_\ph} \cos (\Omega_L \tau)
\sum_{n=1}^N \left(i c_n^\dag c_{n+1} + \hc \right),
\end{equation}
where $\beta$ is the effective electron-photon coupling amplitude and $n_\ph$ is the number of photons in the cavity.
Performing a rotating-wave approximation (RWA), we obtain the time-independent Hamiltonian
\begin{eqnarray}\label{eq:RWA}
H^{\rm RWA} &=& - (\mu_C-\Omega_L) \sum_{n=m_1+1}^{m_2-1} c^\dag_n c_n - \frac{t_C}{2} \sum_{n=m_1+1}^{m_2-2} \left(  c^\dag_n c_{n+1} + \hc \right) \nonumber \\
&& - \frac{\beta \sqrt{n_\ph}}{2} \left[ i \gamma_{m_1} \left(c_{m_1+1} + c^\dag_{m_1 +1}\right) + \gamma_{m_2} \left( c^\dag_{m_2-1} - c_{m_2-1} \right) \right].
\end{eqnarray}
Within this approximation, the effective chemical potential of the central region is shifted to a new value $\mu_C \to \mu_C - \Omega_L$, i.e., the effective gap of the central region can be tuned by changing the microwave frequency $\Omega_L$. Moreover, the coupling strength of MBS to the central region can be controlled by $n_\ph$ which can be changed, for example, by varying the drive strength. The Hamiltonian (\ref{eq:RWA}) has the same structure as the Hamiltonian (\ref{eq:Hsimple}). The resulting splitting between the MBS $\gamma_{m_1}$ and $\gamma_{m_2}$ adjacent to the central region is therefore given by Eq.~(\ref{eq:shift_dyson}) with a shifted chemical potential $\mu_C \to \mu_C - \Omega_L$ and coupling strength $t_{LC} t_{RC} \to \beta^2 n_\ph$. The low-energy dynamics of the MBS is thus governed by a tunable version of the effective Hamiltonian (\ref{eq:H0_low_energy}), and it is possible to perform the rotations (\ref{eq:U}) using microwave pulses.

\begin{figure}[t]
    \includegraphics[width=10cm]{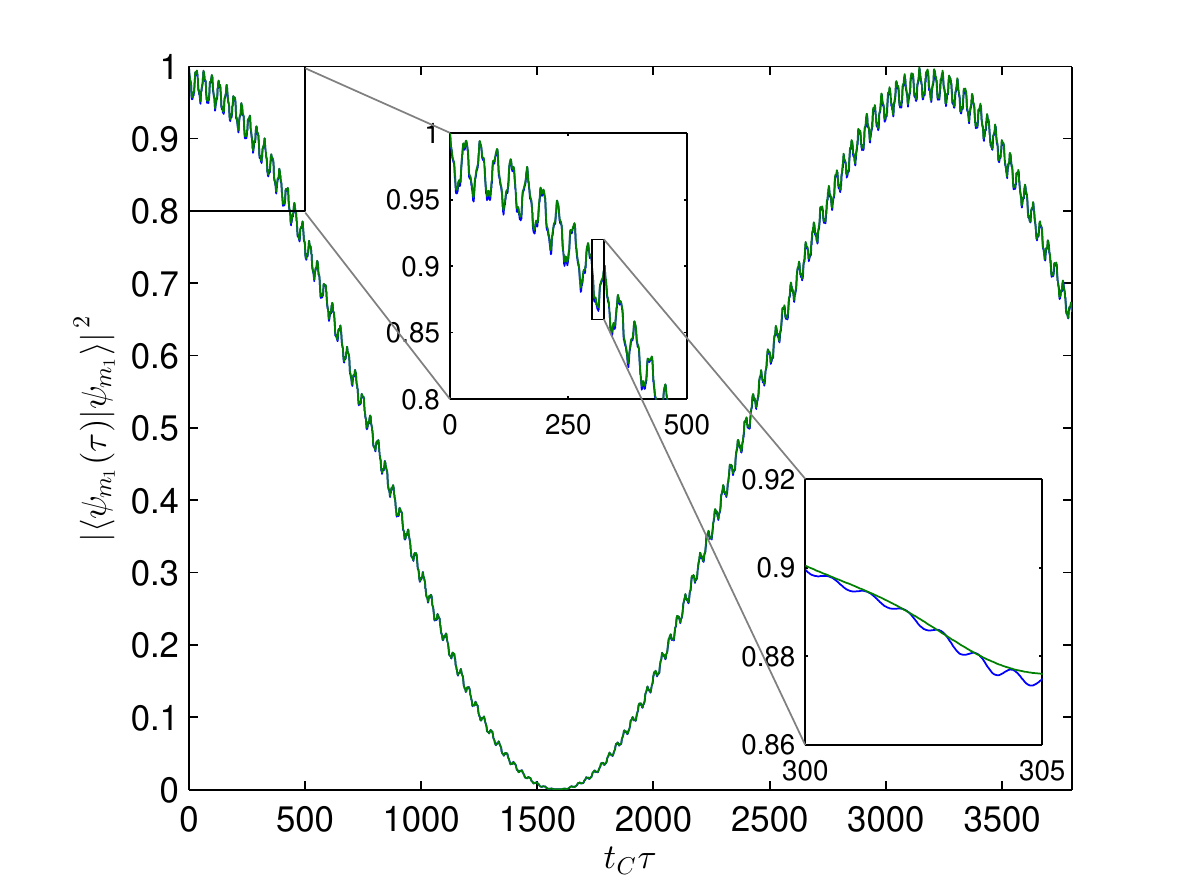}
    \centering
    \caption{Comparison between rotating-wave approximation [green line, Eq.~(\ref{eq:RWA})] and the numerically exact classically driven time evolution [blue line, Eq.~(\ref{eq:Hcl})]. Here, $\tau$ is the time, and $t_C$ the hopping amplitude in the central region. We choose parameters $\Omega_L = 4t_C$ and $\beta = 0.1 t_C$, $N_L = N_R = 2$, $N_C = 4$, $\mu_L = \mu_R = 0$, $\mu_C = 5 t_C$, $\Delta_L = \Delta_R = t_L = t_R = 10 t_C$, $\Delta_C = 0$.}
    \label{fig:ClassicalDrivePlot1}
\end{figure}

We will now discuss the validity of the RWA. For this purpose, we initially prepare the system in the state $\ket{\psi_{m_1}}$. We then start driving at time $\tau = 0$, and observe the resulting Rabi oscillations of the MBS. In Fig.~\ref{fig:ClassicalDrivePlot1}, we plot Rabi oscillations of the MBS $\gamma_{m_1}$ as a function of time, and compare the full numerical solution of the classically driven Hamiltonian (\ref{eq:Hcl}) to the corresponding result of Eq.~(\ref{eq:RWA}). We find that for $t_C \ll \Omega_L < \mu_C$, both results are in excellent agreement. Moreover, the oscillation frequency qualitatively agrees with the approximation (\ref{eq:shift_dyson}). The deviations from the exact numerical result are of the same order as the deviations shown in Fig.~\ref{fig:Splitting} and are mostly due to the fact that the analytical result (\ref{eq:shift_dyson}) assumed $\mu_C \gg t_C$.

\begin{figure}[t]
    \includegraphics[width=10cm]{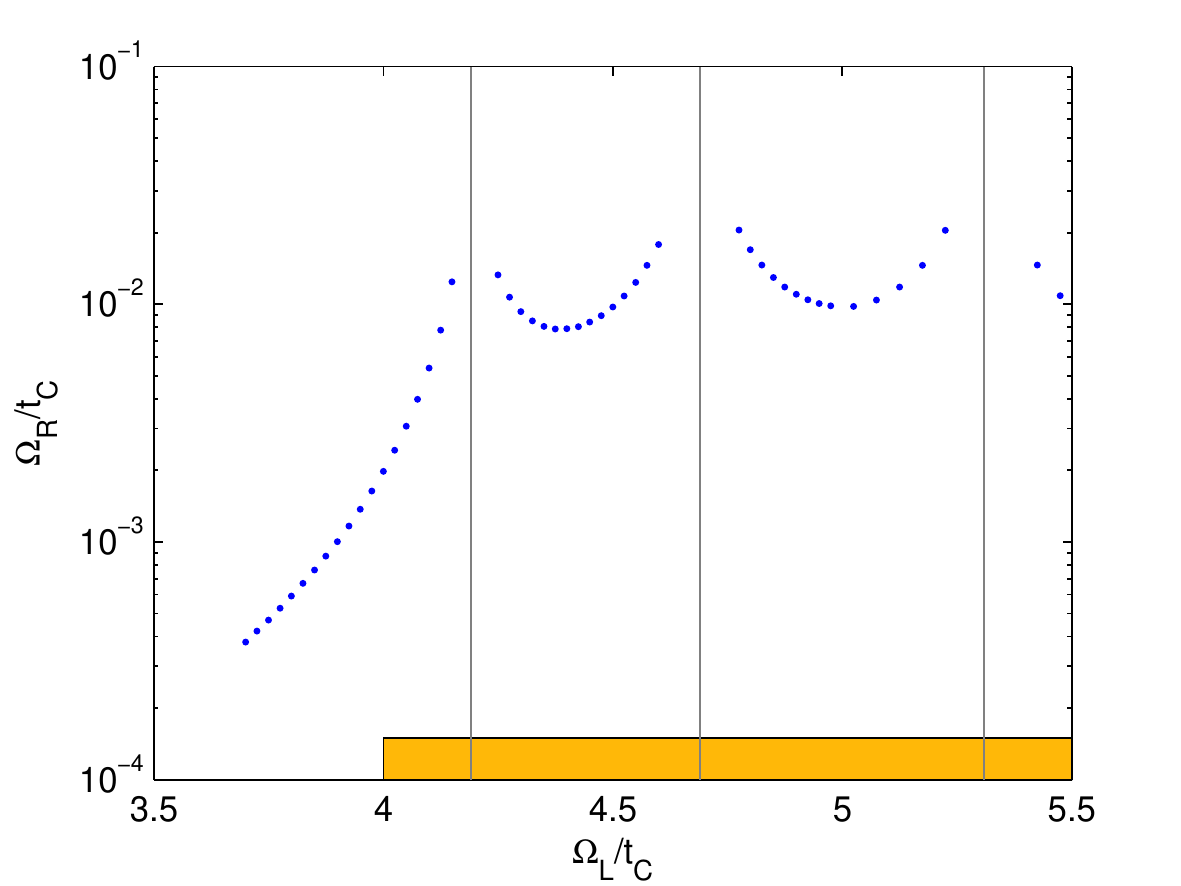}
    \centering
    \caption{Rabi frequency $\Omega_R$ as a function of drive frequency $\Omega_L$ for a classical microwave field. The yellow shaded region shows the position of the conduction band for a continuous central region. The vertical lines denote the eigenenergies of the discrete central region used for the simulation. We choose parameters $\beta = 0.1 t_C$, $N_L = N_R = 2$, $N_C = 4$, $\mu_L = \mu_R = 0$, $\mu_C = 5 t_C$, $\Delta_L = \Delta_R = t_L = t_R = 10 t_C$, and $\Delta_C = 0$.}
    \label{fig:RabiFrequency}
\end{figure}

In Fig.~\ref{fig:RabiFrequency}, we plot the Rabi frequency $\Omega_R$, determined from the numerically exact solution of Eq.~(\ref{eq:Hcl}), as a function of the drive frequency $\Omega_L$. As expected, we find an exponential increase of $\Omega_R$ as $\Omega_L$ approaches the lowest eigenenergies of the central region. In that regime, the exact results agree with the predictions from the RWA. If $\Omega_L$ is close to an eigenenergy, $\Omega_R$ shows resonant behaviour. In this case, the RWA breaks down. In particular at resonance, i.e., if $\Omega_L$ coincides with an eigenenergy of the uncoupled central region, we no longer observe clear Rabi oscillations. For stronger electron-photon coupling, resonances also appear if the drive frequency matches the halves of the eigenenergies (not shown in the figure), signaling two-photon absorption processes which correspond to counter-rotating terms and which are neglected in the RWA.

\subsection{Liouville equation in the quantum regime}
The inhomogeneous Kitaev chain coupled to a damped photon mode can be solved numerically if we truncate the photon Hilbert space to a finite maximum photon number $q_c$. In that case, we can express the system Hamiltonian in the basis $\ket{q,n}$, where $q \in \{0, \ldots, q_c\}$ denotes the number of photons, and the single-fermion states $\ket{n}$ for $n \in \{1, \ldots, 2N\}$ were defined below Eq.~(\ref{eq:psi_j}). The Hamiltonian matrix thus has dimension $2 N q_c$. To find Rabi oscillations numerically, we solve the Liouville equation governing the time evolution of the system density matrix $\rho(\tau)$,
\begin{equation}\label{eq:Liouville}
    \frac{d}{d\tau} \rho(\tau) = \L \rho(\tau).
\end{equation}
The Liouville superoperator $\mathcal{L}$ consists of terms describing the Kitaev chain, the photon field, the electron-photon coupling, and the damping, respectively,
\begin{equation}
    \L \rho = -i[H_K, \rho] - i[H_\ph,\rho] - i[H_{\rm el,ph}, \rho] + \L_{\rm damping} \rho.
\end{equation}
The Hamiltonian $H_K$ for the Kitaev chain is given in
Eq.~(\ref{eq:kitaev}). We again separate the Kitaev chain into
three segments, with boundaries at $1 \leq m_1 < m_2 \leq N$, and
choose the parameters in such a way that the outer segments are in the
topologically nontrivial phase, whereas the central chain is
topologically trivial. Within each of the segments, the parameters
$\mu_n$ and $\Delta_n$ are constant.

The coherently driven cavity mode with resonance frequency $\Omega$ is described by the Hamiltonian $H_\ph$, which contains the photon operators $a$ and $a^\dag$, and a damping term of Lindblad form,
\begin{eqnarray}\label{eq:num_Luncoupled}
    H_{\ph}(\tau) &=& \Omega a^\dag a + a \varphi(\tau) e^{i \Omega_L \tau} + a^\dag \varphi^*(\tau) e^{-i \Omega_L \tau}, \nonumber \\
    \L_{\rm damping} \rho &=& \frac{\kappa}{2} (2 a \rho a^\dag - a^\dag a \rho - \rho a^\dag a).
\end{eqnarray}
Here, $\varphi(\tau)$ encodes the (slowly varying) amplitude and phase of the external drive, $\Omega_L$ is the drive frequency, and $\kappa$ is the photon damping rate. Last but not least, the electron-photon coupling Hamiltonian has the form derived in Eq.~(\ref{eq:map_Helph_lattice}),
\begin{equation}
    H_{\elph} =
    - \beta (a + a^\dag) \sum_{n=1}^{N-1} \left( i c^\dag_n c_{n+1}  + \hc \right)
\end{equation}
which has been used before in Ref.~\cite{trif12}. We solve the Liouville equation by representing the density matrix as a vector with $(2 N q_c)^2$ components, and the Liouville superoperator as a $(2 N q_c)^2 \times (2 N q_c)^2$ matrix acting on this vector. The number of matrix entries scales with the fourth power of the Hilbert space dimension, but the matrix is sparse and thus remains amenable to a numerical solution.

\begin{figure}[t]
    \includegraphics[width=10cm]{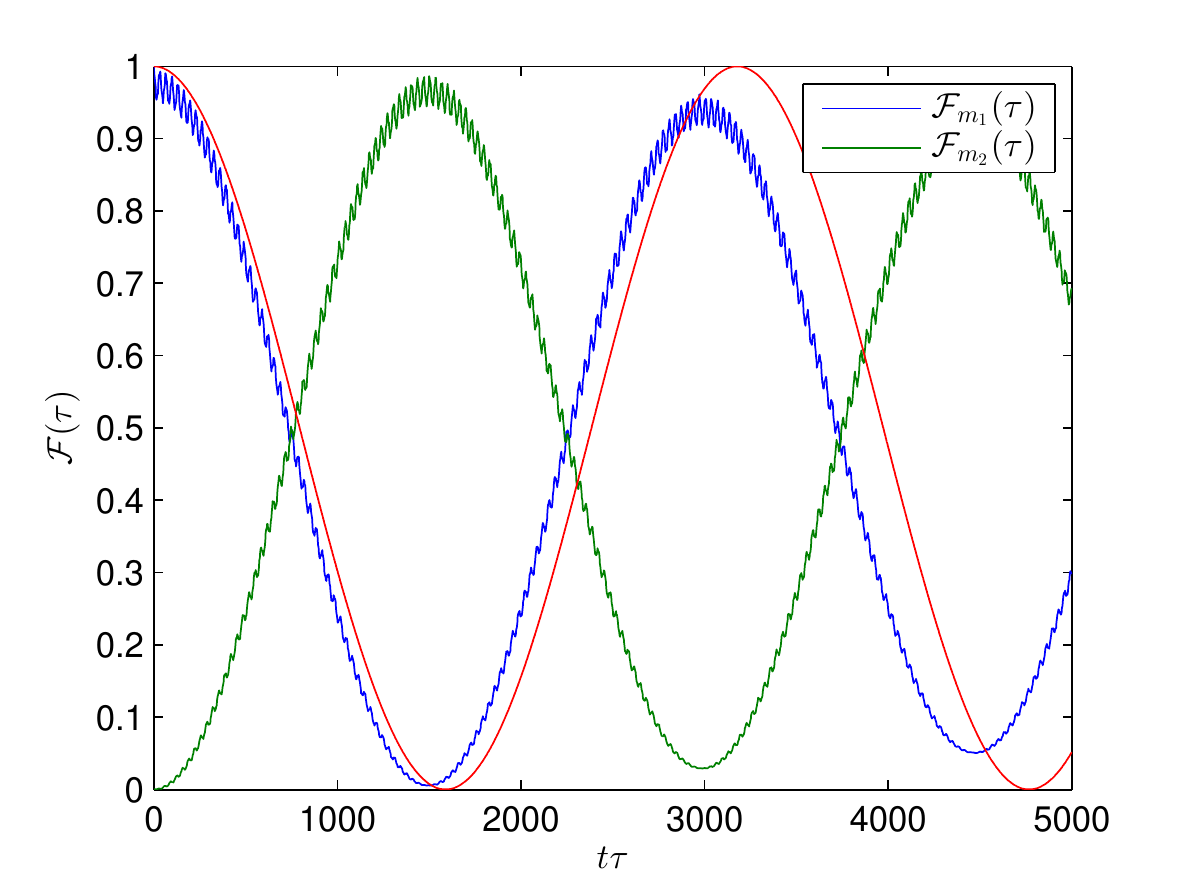}
    \centering
    \caption{Damped Rabi oscillations of the Majorana bound states $\gamma_{m_2}$ (blue line) and $\gamma_{m_3}$ (green line) for a quantized photon field with nonzero linewidth ($\kappa = 10^{-4} t_C)$. The cavity contains one photon in the initial state and is not driven ($\varphi = 0$). The solution was determined from the Liouville equation (\ref{eq:Liouville}). The red line shows the undamped RWA result. We choose parameters: $\Omega = 4 t_C$, $\beta = 0.1 t_C$, $N_L = N_R = 2$, $N_C = 4$, $\mu_L = \mu_R = 0$, $\mu_C = 5 t_C$, $\Delta_L = \Delta_R = t_L = t_R = 10 t_C$, and $\Delta_C = 0$.}
    \label{fig:Liouville}
\end{figure}

The Liouville equation is solved in real time using a fourth-order Runge-Kutta solver. We start with an initial fermionic state $\ket{\psi_{m_1}}$. We set the cavity drive to zero and instead assume that the initial state contains one photon. Therefore, according to Eq.~(\ref{eq:psi_j}), the full system state is $\ket{\psi_i} = \frac{1}{\sqrt{2}} \left( \ket{1,m_1} + \ket{1,m_1+N} \right)$ and the initial density matrix corresponds to the pure state $\rho_i = \ket{\psi_i} \bra{\psi_i}$. To observe Rabi oscillations of MBS, we plot in Fig.~\ref{fig:Liouville} the time-dependent fidelities of the two fermionic states
\begin{equation}
    \mathcal{F}_{m_{1,2}}(\tau) = \Tr \left[ \rho(\tau)\left(  \mathbb{I}_\ph \otimes \ket{\psi_{m_{1,2}}} \bra{\psi_{m_{1,2}}} \right) \right],
\end{equation}
where $\mathbb{I}_\ph$ denotes the identity operator in the photon system. The solution of the Liouville equation reveals damped Rabi oscillations with a damping rate proportional to $\kappa$. For weak damping, the oscillation frequency is slightly reduced compared with the results for classical driving in Sec.~\ref{sec:kitaev_classical}. Moreover, the damping leads to an exponential decay of the amplitude as a function of time.

Within our model, the damping is proportional to $\kappa$ and thus only limited by the photon linewidth. Other conceivable damping mechanisms include fluctuations of the photon field, disorder in the central region, and electron-electron interactions. However, their investigation requires a more detailed study of the actual system hosting the MBS, and is beyond the scope of this paper.

\section{Conclusions}

We have studied an inhomogeneous Kitaev chain
consisting of two topologically nontrivial regions separated by a
topologically trivial, gapped region, and embedded the entire chain in a microwave cavity.
This extension of the Kitaev Hamiltonian has been shown to provide an effective model for a semiconductor nanowire hosting Majorana bound states in the presence of a cavity field. We have presented numerical solutions for the cases of classical and quantized cavity fields valid at arbitrary coupling strengths.
If the microwave frequency approaches the band gap of the topologically trivial region, the coupling between the MBS adjacent to that region is exponentially enhanced. Switching the photon field on and off can then be used to implement controlled rotations of a qubit encoded in Majorana bound states. The qubit rotations achievable using this coupling, combined with braiding operations, are general enough to allow arbitrary single-qubit gates.

\section*{References}

\bibliographystyle{iopart-num}
\bibliography{NJPpaper}

\end{document}